\documentstyle[sprocl,psfrag,graphicx]{article}

\newcommand{\mpl}{m_{\mbox{\tiny pl}}}

\begin{document}

\title{Chaos after Two-Field Inflation}

\author{Richard Easther}

\address{Department of Physics,  Brown University, \\
Providence, RI  02912, USA }

\maketitle\abstracts{We show that two-field inflation can be followed
by an era in which the field dynamics become chaotic, and discuss the
possible consequences of this for two-field inflationary models.}

\section{Introduction}  The simplest realizations of the inflationary
paradigm are based on a single scalar field, minimally coupled to
Einstein gravity. However, there are many incentives for considering
models based on two (or more) interacting fields, including a reduced
need for fine-tuning, and a more diverse range of phenomenological
predictions.\cite{Linde1993a,CopelandET1994a}

It is hardly surprising that more complicated models admit more
complicated behavior. However, the purpose of this paper is to
highlight the possible role of chaotic dynamics in two-field
inflation. In the case of single field, $\phi$, the equations of
motion can be expressed entirely in terms of $\phi$ and $\dot{\phi}$,
and therefore constitute a two dimensional, autonomous system which
does not contain enough degrees of freedom to exhibit chaos. However,
with two scalar fields, the evolution equations form a four
dimensional system, which can be chaotic. Even comparatively simple
chaotic systems may exhibit extremely complex dynamics, and chaotic
behavior differs both qualitatively and quantitatively from
non-chaotic behavior. Consequently, the possible existence of chaos
in two-field inflationary models raises the possibility of new, and
previously unsuspected, phenomenology which may easily be overlooked
by studies based on approximate techniques.

A full description of this work is given in a
preprint,\cite{EastherET1997a} written with my collaborator, Kei-ichi
Maeda.

\section{Chaos and Two-Field Inflation}

The equations of motion for inflationary models driven by two scalar fields
with a combined potential $V(\phi,\psi)$ are
\begin{eqnarray}
&H^2  = \left(\frac{\dot{a}}{a}\right)^2 = \frac{8\pi}{3\mpl^2} \left[
   \frac{\dot{\phi}^2}{2}+ \frac{\dot{\psi}^2}{2} +
V(\phi,\psi)\right]&,\label{Hsqrd}
\\
&\ddot{\phi} = -3H\dot{\phi} -\frac{\partial V}{\partial\phi},
\, \ddot{\psi} = -3H\dot{\psi} -\frac{\partial V}{\partial\psi},
\label{psiddot} &
\end{eqnarray}
where $a$ is the scale factor of the Robertson-Walker spacetime
metric, $\mpl$ is the Planck mass, and dots denote differentiation
with respect to time. The specific results presented here are obtained
for the potential
\begin{equation}
V(\phi,\psi) =
  \left( M^2 - \frac{ \sqrt{\lambda} }{2} \psi^2 \right)^2
  + \frac{m^2}{2} \phi^2 + \frac{\gamma}{2} \phi^2 \psi^2,
  \label{potential}
\end{equation}
but the key ingredient needed for chaos is simply that the fields are
coupled by a term like $\frac{\gamma}{2} \phi^2 \psi^2$, and the
precise form of the potential is not important.

Several workers have examined the possible role of chaotic dynamics in
inflationary models. However, previous studies focussed on the overall
evolution of the universe, and it is not clear that the chaos they
found would lead to consequences detectable by a realistic observer,
whose perspective is confined to the interior of a single universe.

\begin{figure}[tb]
\begin{center}
\begin{tabular}{cc}
\psfrag{xlab}[][]{$\dot{\phi}(0)$}
\psfrag{ylab}[][]{$\dot{\psi}(0)$}
\psfrag{zlab}[][]{$a$}
\includegraphics[scale=.39]{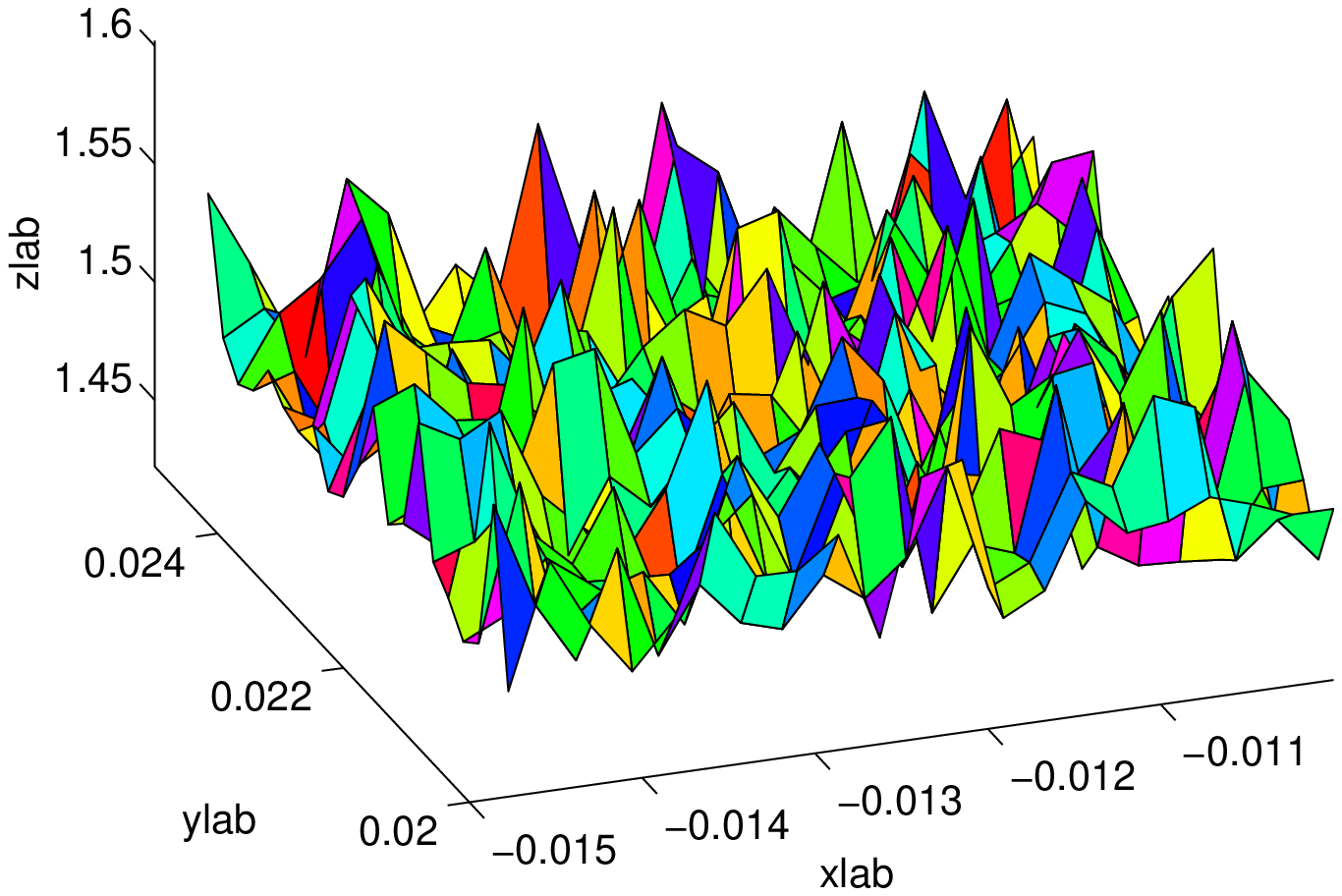}&
\psfrag{xlab}[][]{$\dot{\phi}(0)$}
\psfrag{ylab}[][]{$\dot{\psi}(0)$}
\psfrag{zlab}[][]{$a$}
\includegraphics[scale=.39]{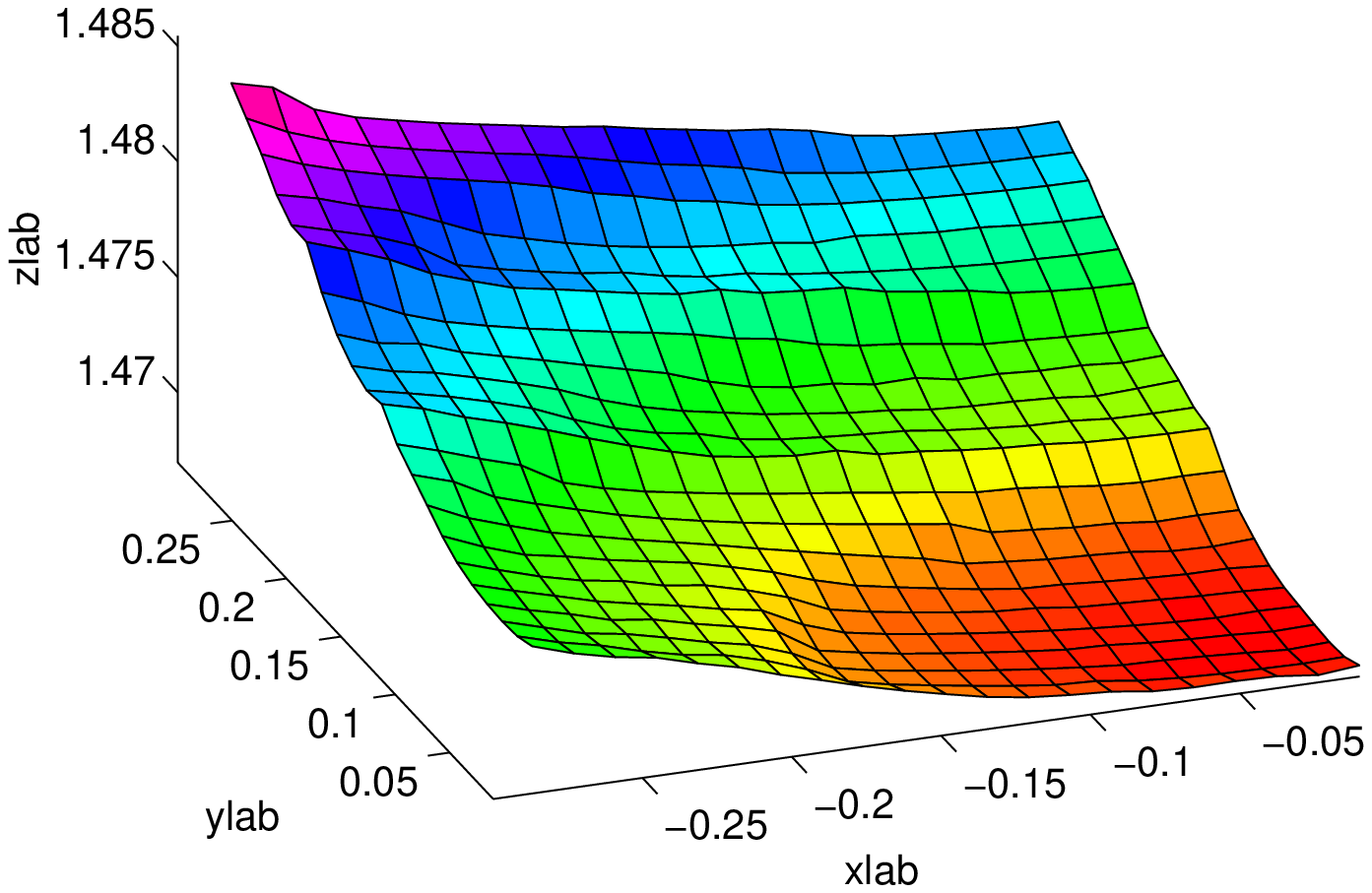}
\end{tabular}
\end{center}
\caption[]{The scale factor at the moment the energy density becomes
equal to $M^4$ and symmetry breaking occurs is plotted as a function
of initial conditions, for $\gamma=0.5$ (left) and $\gamma=0$
(right). The scale factor, $a$, is normalized to be unity at the
beginning of the integration. We chose $\phi(0)=
M\lambda^{1/4}\sqrt{2/\gamma}$, and $\psi(0)=0$, while $\dot{\phi}(0)$
negative so that $\phi$ is initially rolling ``downhill''. The units
are such that $M=m=1$ and $\mpl=5\times10^3$.
\label{crossa}}
\end{figure}

In contrast, we consider a chaotic era which occurs during the
oscillatory phase after the end of inflation, when the damping terms
proportional to $H$ no longer dominate the fields' equations of
motion. The post-inflationary dynamics are then accurately
approximated by the Hamiltonian system that is derived by dropping the
friction term from the equations of the motion. It is straightforward
to show that the frictionless system is chaotic
\cite{EastherET1997a} for a broad range of parameter choices. The next
step is to demonstrate that the chaotic properties persist once the
dissipative effects caused by the expansion of the universe are
reinstated. There is typically a minimum energy (density) below which
chaos cannot occur, so we immediately deduce that the chaotic era in the
post-inflationary universe is transient, since the expansion of the
universe guarantees that the energy density is strictly decreasing.

As a specific example of the consequences of a chaotic era occurring
after the end of inflation, Figure 1 shows how the growth of the
universe during the oscillatory era depends on the precise values of
the fields and their velocities at the end of inflation. We compared a
model with $\gamma=0$, where there is no chance of chaotic evolution,
to one with $\gamma=.5$, which is initially chaotic.  The qualitative
difference between the chaotic and non-chaotic cases is clear. In the
latter case, the amount of growth experienced by the universe after
the end of inflation varies discontinuously with (small) changes in
initial conditions, and the variation is a large fraction of the
average amount of expansion. Conversely, in the non-chaotic case, the
amount of expansion varies smoothly with the changing initial
conditions, and the total amount of variation is much smaller.

\section{Discussion}

We have restricted our attention to ensembles of homogeneous
universes, rather than a single inhomogeneous universe. Since
variations on spatial scales significantly larger than the
post-inflationary horizon volume do not directly affect the dynamics,
we are therefore effectively considering an ensemble of homogeneous
horizon volumes. However, the rapid growth of small initial
differences is the hallmark of chaotic dynamics, so the initial
differences in the field values in different volumes will grow rapidly
during the chaotic era. We expect that the spatial gradient terms in
the fields will become large, and they must be included if we are are
to make quantitative predictions about the role of chaos in the
post-inflationary universe. Work on the inhomogeneous problem is
currently in progress.

\section*{Acknowledgments} The work described here was done in
collaboration with Kei-ichi Maeda. RE is supported by DOE contract
DE-FG0291ER40688 (Task A). Computational work in support of this
research was performed at the Theoretical Physics Computing Facility
at Brown University.

\end{document}